\documentclass[multphys,vecphys]{svmult}

\usepackage{makeidx}     
\usepackage{graphicx}    
\usepackage{multicol}    

\makeindex             

\begin{document}

\title*{Supernova Spectra}
\author{Massimo Turatto\inst{1}}
\institute{Osservatorio Astronomico di Padova -- INAF, 
vicolo dell'Osservatorio 5, 35122 Padova, Italy
\texttt{turatto@pd.astro.it}}
%
%
\maketitle

\abstract In this paper are summarized the main advances of the last  years in the
field of SN spectra . The arguments against a monodimensional sequence for SNIa are
discussed as well as the efforts to improve the temporal and spectral coverage of
this kind of SNe, with the aim to understand the physics of the explosions for
their use as cosmological distance indicators. Although variety is the main
character of core--collapse SNe, we have been recently surprised by both
exceptionally under-- and over--energetic explosions. The main properties of these
two extreme subclasses are here reviewed.

\section{Introduction}
\label{sect:intro}

Spectra provide most of the physical information on Supernovae. 
Their early analysis has shown 
a variety of forms and evolutions.
Indeed the identification of different spectral lines 
reveals the presence of several different ions in the
layers above the photosphere, suggesting the existence of various 
progenitors and explosion models.
Spectra also allow us the direct determination of the physical conditions of the
emitting regions, while the line profiles provide the kinematics.

Supernova spectra evolve rapidly: the effective temperature and expansion velocity
decrease, and the spectral lines change. Indeed, because of the expansion, the
photosphere recedes into the ejecta and different layers are progressively exposed.
In this way the analysis of SN spectra taken at different epochs allows us, at
least in principle, to make the tomography of the exploding stars and to
reconstruct their entire structures. A major limitation is that spectral features
in SN spectra  are generally blended due to the large expansion velocities and the
full information can be extracted only with the use of complex spectral
modeling which try to deal in a consistent way the luminosity, abundances,
stratification, temperature, velocity and time evolution.

In general, the light from SNe is travelling for several megaparsecs before
reaching the observers thus contains information on the circumstellar,
interstellar and intergalactic matter it has passed through. 
In principle, high resolution spectroscopy might allow us to determine 
the distribution and physical conditions of the intervening medium and, in turn,
the total reddening and extinction suffered by the light. 

Finally, in the recent years spectral observations on faint, distant  
of SNIa have provided sufficiently accurate redshifts which, once coupled with the
photometry, have inspired a new vision of the Universe we are living in.

This review discusses a number of hot topics concerning SN spectra while
reference is given to other reviews for the detailed description of 
the properties and evolution
of various spectral types \cite{fil97, tur03,wheben00}.

\section{A spectral sequence for SNIa ?}
\label{sect:snia}

SNIa explode in all types of galaxies, in ellipticals  as well as in spirals,
but in the latters are not closely associated with the spiral arms as other SN
types. The spectra are characterized by lines of intermediate mass elements
such as Ca, O, Si and S during the peak phase, and by the absence of H at any
time. With age the contribution of the Fe lines increases and several months
past maximum the spectra are dominated by [Fe~II] and [Fe~III] lines. The
overall homogeneous spectroscopic and photometric behavior has led to a
general consensus that they are associated with the thermonuclear explosions of
white dwarfs \cite{bran95}.

During the last decade a new scenario for the SNIa has been progressively
developed. In particular, a correlation between the peak luminosity and the
shape of the early light curve was found, with brighter objects having a  rate
of decline slower than dimmer ones \cite{phil93,perl97,phil99,riess98}. An
analogous spectroscopic sequence has been found \cite{nugent}, according to
which the absolute magnitude of SNIa and, in turn, the rate of decline, is
correlated to $\cal R$(SiII), the ratio of the depths of two absorption
features at 5800 and 6100 \AA, usually attributed to Si~II. Synthetic spectra
modeling indicates that most of the spectral differences are caused by
variations in the effective temperatures, likely due to different amounts of
$^{56}$Ni produced in the explosions. The finding that fainter SNe show slow
expansion velocities both at early \cite{branchvdb} and later epochs 
\cite{tur91bg} is consistent with such scenario.

In first approximation all SNIa can be accommodated into such one--parameter
sequence which can be regarded as a sequence of explosion strengths. Within
such scheme fit the bright, slowly declining SN~1991T, which
did not exhibit Si~II or Ca~II absorption lines in the premaximum spectra but
had a normal behavior starting one month after maximum, and the faint,
intrinsically red and fast declining SN~1991bg, which showed a slow expansion
velocities and a deep trough around 4200 \AA\/ produced by Ti~II.

In the last years, however, new findings have challenged such
a monoparametric sequence. 
Hatano et al. \cite{hata00} have shown that the $\cal R$(SiII)
does not correlated with the photospheric velocity deduced from 
the Si~II $\lambda6355$ absorption, 
as one would expect, and propose the existence of two or more 
explosion mechanisms as possible explanation
for the lack of correlation.
Also, a recent reanalysis \cite{bene03} has shown that
while the $\cal R$(SiII) vs. $\Delta m_{15}$  relation holds 
for $\Delta m_{15} \ge 1.2$, at smaller values it breaks down,
thus questioning the correspondence between spectroscopic and photometric
parameters.

A number of objects have now good sequences of spectra, some starting very
early after the explosion, which make possible to study the temporal 
evolution of $\cal R$(SiII). It is found that before maximum $\cal R$(SiII)
exhibits a dramatic evolution with
opposite trends for various objects \cite{bene03}.

\begin{figure}
\centering
\includegraphics[height=9cm]{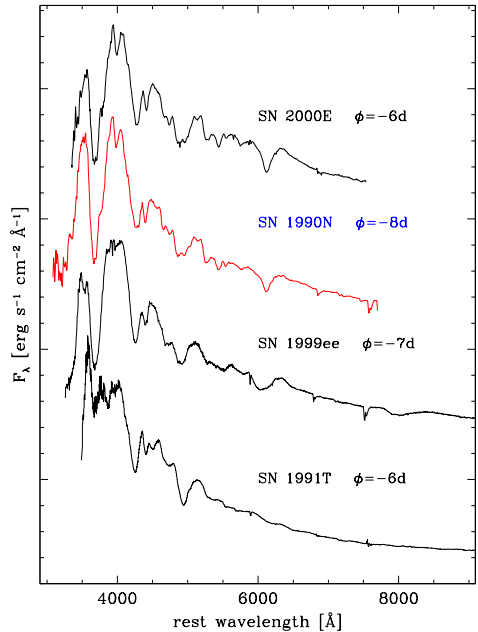}
\label{fig:cfr_fm7}
%
\caption{Comparison of the spectra of SNe 2000E ($\Delta m_{15}=0.94$
mag (100d)$^{-1}$),
1990N ($\Delta m_{15}=1.05$), 1999ee ($\Delta m_{15}=0.92$),
1991T ($\Delta m_{15}=0.94$), taken about 1 week before 
maximum light. They have been corrected for extinction and reported to the
parent galaxy rest frame.}
\end{figure}

Moreover, objects with similar decline rates can show different spectral
features, especially before maximum. In Figure \ref{fig:cfr_fm7} the spectra of four objects taken about
one week before B maximum are compared. The spectrum of SN~2000E 
($\Delta m_{15}=0.94$ mag (100d)$^{-1}$) shows noticeable differences with respect
the two SNe at the bottom, having the same decline rates, but is very similar to
that of SN~1990N ($\Delta m_{15}=1.05$). 
Not only it differs from that of SN~
1991T ($\Delta m_{15}=0.94$) which is well known for not displaying at this
epoch the characteristic Si~II line but shows disturbing differences with the
normal SN~1999ee ($\Delta m_{15}=0.92$). In particular, the profile of the 
Si~II 6355$\lambda$ line is definitely broader in SN~1999ee indicating the
presence of Si~II at high velocity. Also the SiII 4130$\lambda$ line, 
which can be used for classifying
high redshift SNe as type Ia, is nearly washed out. Note also that the
analysis of the Ca~II IR triplet of SN~1999ee before
maximum  (Mazzali priv. communication) requires high velocity material in analogy to what found for 2001el
\cite{thomas03}.

Two other odd SNIa are 2000cx and 2002cx which
do not fit into any photometric and spectroscopic 
sequence still showing the main characteristics of type Ia 
SNe \cite{li01a, li03}.


\section{Trends in SNIa spectroscopy}

More and better data are required to clarify which parameters
govern the SNIa explosions and to validate the proposed progenitor scenarios.
To solve these questions large collaborations have born 
which have as immediate goal the intensive, multiwavelength monitoring
of nearby SNIa.

Examples of such intensive monitoring are the studies of SNe 1999ee \cite{mario99ee}
and 2002bo \cite{bene03}, for which the rise to maximum
has been sampled daily starting about two weeks before maximum. The
analysis of these spectra has allowed the discovery of strong time evolutions 
of the ratio $\cal R$(SiII) with individual behaviors for each SNIa
(cfr. Fig. 9 of Benetti et al. \cite{bene03}), as mentioned above.

These two objects have been extensively monitored spectroscopically also in the
infrared. The comparative analysis with the IR spectra of SN~1994D
\cite{meikle96} at about day $-8.5$, +11.1 and +29.4 confirms that the overall
spectroscopic homogeneity among normal SNe~Ia extends to the IR-domain, with
small variations.  Spectral synthesis on the spectra of SN~2002bo favors the
MgII identification for the 10800\AA\ line and that of SiII 16910\AA~ and MgII
16760/800 \AA (with the SiII dominant) for the broad P-Cygni profile observed
at $\sim$16000\AA. On the contrary, major
differences are visible when the comparison is made with peculiar objects, e.g.
the faint SN 1999by \cite{hofl02}, especially because of the presence of strong
C~I  and O~I lines.

The optical and the IR spectra of SNIa \cite{mario99ee} shows different
behaviors: a) while the optical spectrum before maximum is dominated by strong
P Cygni profiles of intermediate-mass elements, such as Ca II, Si II, Mg II, S
II, the IR is characterized by a smooth, almost featureless continuum; b) the
lines of iron group elements, such as Co II, Fe II and Ni II, emerge in the IR
as soon as one week past maximum, definitely earlier than in the visual. This
supports the suggestion \cite{spyro94} that the IR photosphere
recedes rapidly to the center of the supernova, while at optical wavelengths the
greater opacity arising from the higher spectral density of lines keeps the
photosphere at higher velocities. At the longer wavelengths the overall emission
increases after maximum light as a result of a shift in ionization to lower
ionization species, which have greater emissivity in the near IR.  It appears,
therefore, that the J-band deficit is due to the relative absence of lines in
the 12000 Å region, rather than increasing opacity, and that the secondary
maxima exhibited by the IR light curves are due to the increasing release of
energy through lower optical depth IR transitions. The prominent postmaximum
emission features displayed by SNe 1999ee and 2002bo in the H and K bands lend
support to this scenario \cite{mario99ee}.

A new powerful tool for understanding the nature of SNIa is spectropolarimetry. 
SNIa polarization is usually very small, hence difficult to measure.
In most cases only upper limits have been
provided, in others polarization of the order of
0.2--0.3\% (corresponding to an asphericity of $\sim$10\%) has been
detected before maximum light, likely due to a distorted photosphere
or element distribution.  Such asphericity may cause a directional
dependance of the luminosity and a corresponding
dispersion in the brightness-decline relation of SNe Ia.  Although this
may not jeopardize the use of SNIa as distance indicators, it might
intrinsically limit the accuracy reachable via SNIa.

Very interesting spectropolarimetric data have been
collected for special SNIa.  High polarization (0.7\%) was found in 
the subluminous SN 1999by \cite{how01}, which may
suggest a relation between the observed asymmetry and the mechanism
that produces this kind of underenergetic SNIa.  The high velocity
components (v$\sim 25000$ km/s) of the CaII IR triplet observed in
SN~2001el have shown a polarization of about 0.7\%, much higher than
the continuum (0.2\%) indicating that kinematically and geometrically
distinct features can exist in SNIa \cite{wang}.  
High signal--to--noise data for more objects are definitely needed in order
to understand if these asymmetries are the rule in SNIa, and to get insights on the
the geometry of the precursor systems.

A special case is that of SN~2002ic, the first SNIa for which H has been
unequivocally detected \cite{hamuy03}. Indeed both the light curve and the
spectral line appearance points toward a strong interaction of the
SN ejecta with a dense CSM. At the early phase the characteristic spectrum of a
SNIa seemed veiled by a strong continuum. With time the spectrum evolved to
resemble those of SNe 1997cy and 1999E, which are commonly considered as SNIIn
\cite{germ, tur97cy, rigon}. Also the profile of the H$\alpha$ emission requires
at least two components with different widths to provide a reasonable fit. It
has, therefore, suggested that some SNIIn are the outcome of thermonuclear
explosions rather than core collapses. Despite the early claim that these
observations were a proof that the progenitor system 
of SN~2002ic was a binary consisting of
a C/O WD and an AGB star \cite{hamuy03}, it may well be that it is the result of
the merging of a WD with the core of an AGB star, and that the H we observe was
previously ejected during a common envelope phase \cite{livio03}. The rarity of
such an event and the large amount of H required favors the latter hypothesis. A
polarization of $p \ge1$\% \cite{kawa} has been detected also in SN~2002ic.

\begin{figure*}
\centering
{\includegraphics[height=13cm,angle=-90]{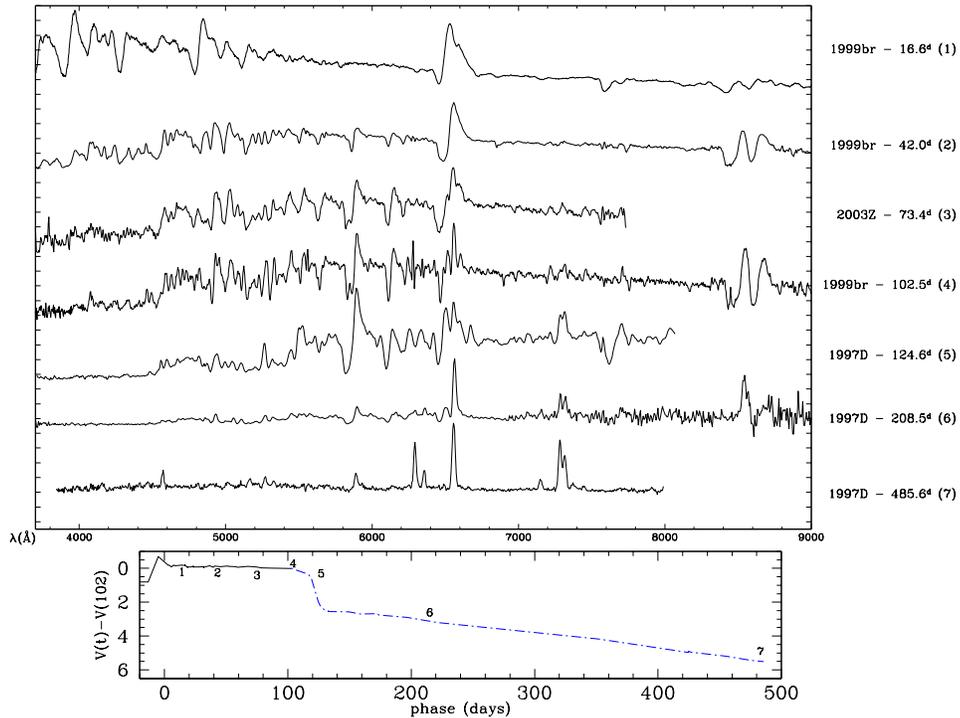}}
\label{fig:faint}
\caption{Spectral evolution of low--luminosity SNII. The objects
and the days after the estimated epochs of explosions are marked on the right.
In the bottom panel are marked the corresponding epochs on a
schematic light curve.}
\end{figure*}

\section{Core Collapse Supernovae}
\label{sect:ccsn}

Stars of initial mass larger than about 8 M$_\odot$ undergo the collapse of
the core after burning H, He, C, Ne, O, Si. From such collapse a SN usually
results.

Core collapse SNe  can vary considerably as luminosity, spectral lines and time
evolution. The main parameter governing the observed
diversity is the envelope mass \cite{nomo95}. Stars retaining H will display the
lines of this element during their entire evolution and are called SNII. If the
H mass is large ($10 - 15$ M$_\odot$) the release of the energy deposited by
the shock in the envelope and by the recombination can sustain the luminosity
for months, producing the so-called SNII Plateau. If the envelope is less
massive ($1 - 2$ M$_\odot$), stripped by a companion or lost by stellar wind, 
the SN quickly fades producing the so--called SNII Linear.

In both cases the spectrum evolves from a very blue, featureless continuum short
after the burst (color temperature higher than 10$^4$ $^\circ$K) to one dominated 
by H and He in a few days. Then, as the temperature continues to fall, other
low excitation lines of Na~I, Ca~II, Fe~II, Sc~II appear, all with P--Cygni
profiles. Subsequently, as the entire star becomes transparent and the light curve
settles onto the radioactive tail, the spectrum enters the nebular phase, dominated
by H$\alpha$ and forbidden emission lines of [O~I], [Ca~II], [Fe~II] and Mg~I].

After SN 1987A, the best studied (but somehow atypical) object,  whose detailed
spectral observations provided evidences of clumps and mixing in the ejecta 
\cite{hanusch87, spyro93} as well as of dust formation \cite{danz91}, other 
SNII have been studied extensively. An interesting case is SN~1999em 
\cite{baron00, hamuy01, leon01, abu03}. The analysis of the H$\alpha$ structure at photospheric epoch
revealed non spherical ejection of $^{56}$Ni, while the transformation of the
[O~I] 6300 $\lambda$ line profile around day 500 showed that dust formed earlier
and at lower velocity than in SN~1987A, probably because of the lower temperature
due to a smaller amount of $^{56}$Ni ejected. It appears, therefore, that these
phenomena are common in core collapse SNe, though with distinctive characteristics
for each object.

Core--collapse SNe which have lost most of their H and even most, or all, 
their He envelope, are called SNIb and Ic, respectively. Indeed the transition
between SNII and SNIb, and between SNIb and SNIc is not sharp. SN1993J,
celebrated with this Conference, is the best example (but not the only one) of a
SN transforming with age from a type II to a type Ib SN, i.e. from one dominated by H to
one dominated by He, with only residual H lines. For this reason it is 
called of type IIb \cite{fil_math}. 

At a deep scrutiny, H has been found in the spectra of other SNIb.
SN~2000H is a remarkable case, but Branch et al. 
\cite{bran02} have shown that tiny differences in the H mass can vary the optical
depths of H lines and make the transition from type IIb to type Ib SNe.
Similarly, an increasing number of SNIc show evidence of He, e.g.
SN~1990B, SN~1987M and even the prototypical SN1994I  \cite{fil95, cloc96}.
Again, it appears that moderate difference in the He mass can explain the
spectroscopic difference between typical SNIb and SNIc \cite{bran02}. Different
is the case of SN~1999cq, in which He with expansion velocity much slower than
other lines, points to the interaction of the ejecta with a dense shell of
almost pure He originating from stellar wind or mass transfer to a companion
\cite{mat00}.

\subsection{News on Core Collapse SN spectra}
\label{sect:snii}

%

In the last years new light on the lower end of the luminosity function of SNII
has been  shed \cite{zamp_d_br, pasto03}. SN1997D showed from the discovery
unprecedented properties. The expansion velocity deduced from the
displacement of the absorption was extremely low  and the continuum
intrinsically red. Moreover, the  luminosity was very low both at discovery and
in the nebular stage, indicating the ejection of a very little amount of 
$^{56}$Ni \cite{tur97d}.

Four new objects with similar properties have been recently added 
\cite{pasto03}. Although the observations of individual objects are erratic and
incomplete, because hampered by the faintness of the SNe, taken together these
data show a common evolutionary path. Figure \ref{fig:faint} summarizes the
spectroscopic evolution: during the first 50 days (spectra 1 and 2) the spectra
change from a continuum dominated by Balmer lines to a more complex appearance
with strong Na~I, Ba~II and Ca~II. The expansion velocities, lower than in normal
SNII, decrease from 5000 to 3000 km s$^{-1}$. During the second half of the
plateau (spectra 3 and 4) these peculiarities strengthen
significantly: the absorption troughs move to even lower velocities, down to
1000--1500 km s$^{-1}$, the continuum becomes redder and low excitation lines
of Ba~II, Sc~II, Fe~II, Sr~II and Ti~II appear. In particular, the Ba~II lines become
the strongest features of the entire spectrum, even stronger than H$\alpha$.
During the postplateau decline (spectrum 5) the transition to the 
nebular phase begins and forbidden lines (e.g. [Ca~II] 7291--7323 \AA) emerge.
The spectra of the latest epochs (6 and 7) resemble those of normal SNII, with the usual (though
narrower) emissions of H$\alpha$, Na~I, [O~I], [Ca~II], [Fe~II], Mg~I]. 

Pastorello et al.  \cite{pasto03} conclude that low--luminosity CC--SNe are
similar to typical CC--SNe in having a clear plateau, lasting for $\sim$100~days,
followed by a late--time decline driven by the decay of $^{56}$Co, and typical
spectral lines at all phases.  However, they keep distinctive characteristics in
that (a) during the plateau phase the luminosity is at least a factor 10 times
less than  in typical CC--SNe, (b) the expansion velocity is unusually low at all
epochs, and (c) the mass of $^{56}$Co which drives the late--time tail is at least
a factor $\sim$10 lower than normal.

Important advances have been obtained also at the opposite extreme, i.e. for
luminous and energetic SNe, often called hypernovae. In particular, large
interest have received a number of SNe associated to GRBs. The first and best
studied case is SN 1998bw which coincided in time and space with GRB980425 
\cite{gal98} and was a powerful radio and X--ray emitter. In the optical most of its
peculiarity stayed in the unprecedented broadness of the spectral features
corresponding to expansion velocities of the ejecta as high as $3 \times 10^4$
km s$^{-1}$ at maximum light \cite{patat98bw}. The high kinetic energy together
with the high luminosity indicated an explosion energy of about
$5 \times 10^{52}$ ergs, if spherical symmetry is assumed \cite{iwa98}. Detailed spectral
modeling \cite{iwa98, bran01a, maz01, iwa03} has shown that the apparent
emissions, which at maximum peaked around 5000, 6300 and 8500 \AA, were actually
low opacity regions of the spectra from which photons could escape. Lines of Si~
II, O~I, Ca~II and FeII have been identified, as well as He~I 10830 \AA.

Only during the nebular phase SN~1998bw reentered into the 
conventional taxonomical scheme with [O I], Mg I], [Fe~II], [Ca~II] emissions
which confirmed the early hypothesis that it was a peculiar case of SNIc.
Also in the nebular phase the lines were unusually broad
($9800\pm500$ km s$^{-1}$ on day 201 \cite{patat98bw}).

After SN~1998bw several other hypernovae have been recognized, some were SNIc
bearing strong resemblance to SN~1998bw but with smaller KE (1997ef, 1997dq, 1999as, 2002ap \cite{
iwa00, mat01, iwa03, maz02}), other were SNIIn (1997cy and 1999E \cite{germ,
tur97cy, rigon}), which displayed narrow H$\alpha$ on the top of broad wings and
broad light curves, clear indications of ejecta--CSM interaction. For some of
them the possible association with GRB has been claimed but no firm conclusion was
reached. Other cases of possible SN--GRB association have been reported on the
basis of 'bumps' detected in the light curves of GRB afterglows. The early
spectroscopy of these bumps was intriguing but not conclusive 
\cite{garna03, mdv03}. 

Excitement mounted when, short before this Conference, spectra 
of the rebrightening of the afterglow of the nearby GRB030329
(z=0.1685) were secured \cite{stanek, mat03}. Already one week after the burst a SN
spectrum, with emissions at approximately 5000 and 4200 \AA, was detected 
and it dominated over the power-law continuum few days later (day $\sim11$). 
This SN, named 2003dh, 
resembled, both as spectral evolution and light curve, SN~1998bw at 
the corresponding epoch from the associated GRB.
A more accurate analysis and spectral modeling suggested 
that SN~2003dh was intermediate between
SN~1998bw and SN~1997ef, as to kinetic energy and $^{56}$Ni production
\cite{maz03} and pointed out the need for asymmetric explosions.
Whatever the physics of the explosion, this new example 
provides solid evidence that at least some GRBs arise from core-collapse SNe
and opens new frontiers in the SN research.

%
%

\bigskip
ACKNOWLEDGMENTS.
This research is supported in part by the European Community's Human Potential 
Programme under contract HPRN-CT-2002-00303, and grant Cofin 2001021149
of the Italian Ministry of Education, University and Research.

%
%
%
%
%
%

\begin{thebibliography}{99.}
%
%
\bibitem{ben97d} S.~Benetti, et al: 2001, MNRAS \textbf{322}, 361
\bibitem{bene03} S., Benetti, et al.: 2003, MNRAS in press astro-ph/0309665
\bibitem{baron00} E., Baron, et al.: 2000, ApJ \textbf{545}, 444,
\bibitem{branchvdb} D.~Branch, S. Van Den Bergh: 1993, AJ \textbf{105} 2231
\bibitem{bran01a} D.~Branch: In \textit{Supernovae and Gamma--Ray Bursts}
      eds. M. Livio, N. Panagia (Cambridge University Press, Cambridge 2001), p.96
\bibitem{bran95} D.~Branch, et al.: PASP \textbf{107} 1019 (1995) 
\bibitem{bran02} D.~Branch, et al.: 2002, ApJ \textbf{566}, 1005
\bibitem{cloc96} A. Clocchiatti, et al.: 1996, ApJ \textbf{462} 462
\bibitem{danz91} I.J.~Danziger, et al.: In: \textbf{Supernova 1987A and other supernovae}, 
   ed. I. J. Danziger, K. Kjar (ESO Conference and Workshop Proceedings) p.217
\bibitem{mdv03}  M.~Della Valle, et al.: 2003, A\&A \textbf{406} 33
\bibitem{abu03} A., Elmhamdi, et al.: 2003, MNRAS \textbf{338}, 939
\bibitem{fil97} A.V., Filippenko: 1997, ARAA \textbf{35}, 309
\bibitem{fil_math} A.V., Filippenko, T., Matheson: Optical, Ultraviolet, and Infrared
     Observations of SN 1993J. In: \textit{Supernovae: 10 Years of 1993J} ed. J. Marcaide,
     K. Weiler (2004) pp.
\bibitem{fil95} A.V. Filippenko, et al: 1995, ApJ \textbf{450} L11
\bibitem{gal98} T.J.~Galama, et al.: 1998, Nature \textbf{395} 670
\bibitem{garna03} P.M.~Garnavich, et al.: 2003, ApJ \textbf{582} 924
\bibitem{germ} L. Germany, et al.: 2000, ApJ \textbf{533}, 320
\bibitem{hamuy01} M.~Hamuy, et al.: 2001, ApJ \textbf{558}, 615
\bibitem{mario99ee} M.~Hamuy, et al.: 2002, AJ \textbf{124} 417
\bibitem{hamuy03} M.~Hamuy, et al.: 2003, Nature \textbf{424}, 651

\bibitem{hanusch87} R.W.~Hanuschik, et al.: 1988, MNRAS \textbf{243} 41 
\bibitem{hata00} K.~Hatano, et al.: 2000, ApJ \textbf{543}, L49
\bibitem{hofl02} P.~H\"oflich, et al.: 2002, ApJ, \textbf{568} 791
\bibitem{how01} D.A.~Howell, et al.: 2001, ApJ \textbf{556} 302
\bibitem{iwa98} K. Iwamoto, et al.: 1998, Nature \textbf{395}, 672
\bibitem{iwa00} K. Iwamoto, et al.: 2000, ApJ \textbf{534}, 660
\bibitem{iwa03} K. Iwamoto, et al.: SN~1998bw and Hypernovae.  In: \textbf{Supernovae 
     and Gamma--Ray Bursters}, Lecture Notes in Physics, ed. K.W. Weiler (Springer, Berlin 2003), pp 243, 281
\bibitem{kawa} K.S., Kawabata, et al.: 2003, IAU Circular 8161
\bibitem{leon01} D.C., Leonard, et al.: 2002, PASP \textbf{114}, 35
\bibitem{li01a} W. Li, A.V., Filippenko, E., Gates, et al.: 2001, PASP \textbf{113}, 1178
         ApJ \textbf{546}, 734
\bibitem{li03} W. Li,W., A.V., Filippenko, R., Chornock, et al.: 2003, astro-ph/0301428 
\bibitem{livio03} M. Livio, A.G., Riess: 2003, ApJ \textbf{594}, L93
\bibitem{mat00} T.~Matheson, et al.: 2000, AJ \textbf{119} 2303
\bibitem{mat01} T.~Matheson, et al.: 2001, AJ \textbf{121} 1648
\bibitem{mat03} T.~Matheson, et al.: 2003, ApJ submitted (astro-ph/0307435)
\bibitem{maz01} P.~Mazzali, et al.: 2001, ApJ \textbf{559} 1047
\bibitem{maz02} P.~Mazzali, et al.: 2002, ApJ \textbf{572} L61
\bibitem{maz03} P.~Mazzali, et al.: 2003, ApJ submitted (astro-ph/0309555)
\bibitem{meikle96} W.P.S.~Meikle, et al.: 1996, MNRAS \textbf{281} 263
\bibitem{nomo95} K. Nomoto, K., Iwamoto, T., Suzuki: 1995 Phys. Rep. \textbf{256}, 173
\bibitem{nugent} P.~Nugent, et al.: ApJ \textbf{455} L147 (1995)
\bibitem{pasto03} A., Pastorello, et al.: 2003, MNRAS in press (astro-ph/0309264)
\bibitem{patat98bw} F.~Patat, et al. : 2001, ApJ \textbf{555} 917
\bibitem{perl97} S.~Perlmutter et al: 1997, ApJ \textbf{483}, 565
\bibitem{phil93} M.M.~Phillips: 1993, ApJ \textbf{413} L105  
\bibitem{phil99} M.M.~Phillips: 1999, AJ \textbf{118} 1766
\bibitem{riess98} A,G.~Riess, et al.: 1998, AJ \textbf{116} 1009
\bibitem{rigon} L. Rigon, et al.: 2003, MNRAS, \textbf{340}, 19
\bibitem{schmidt} B.P. Schmidt, et al.: 1994, AJ \textbf{107}, 1444
\bibitem{spyro93} J.~Spyromilio, et al.: 1993, MNRAS \textbf{263} 530     
\bibitem{spyro94} J. Spyromilio, et al.: 1994, MNRAS \textbf{266} L17  
\bibitem{stanek} K.Z.~Stanek, et al.: 2003, ApJ \textbf{591} 17
\bibitem{thomas03} R.C. Thomas, et al.: 2003, ApJ in press (astro-ph/0302260) 
\bibitem{tur91bg} M.~Turatto, et al: 1996, MNRAS \textbf{283} 1
\bibitem{tur97cy} M.~Turatto, et al: 2000, ApJ \textbf{534} L57
\bibitem{tur97d} M.~Turatto, et al: 1998, ApJ \textbf{498}, L129
\bibitem{tur03} M. Turatto: Classification of Supernovae. In: \textit{Supernovae 
     and Gamma--Ray Bursters}, Lecture Notes in Physics, ed. K.W. Weiler (Springer, Berlin 2003), pp 21--36 
\bibitem{wheben00} J.C.~Wheeler, S.~Benetti: In: \textit{Allen's Astrophysical Quantities}, 
         ed.~by A.N.~Cox (Springer, New York 2000) p.~451
\bibitem{wang} L. Wang, et al.: 2003, ApJ \textbf{591}, 1110
\bibitem{zamp_d_br} L. Zampieri, et al.: 2003, MNRAS \textbf{338}, 711
%




\end{thebibliography}
%



\printindex
\end{document}